\documentclass[cameraready]{Interspeech}

\usepackage{graphicx} 
\usepackage{amsmath,amsthm}
\usepackage{enumitem}
\usepackage{comment}
\usepackage{xspace}
\usepackage{etoolbox}
\usepackage{graphicx}
\usepackage{algpseudocode}
\usepackage{algorithm}

\newtheorem{definition}{Definition}
\newtheorem{theorem}{Theorem}
\newtheorem{assumption}{Assumption}

\algrenewcommand\algorithmicrequire{\textbf{Input:}}
\algrenewcommand\algorithmicensure{\textbf{Output:}}

\newcommand{\encode}{\phi}
\newcommand{\decode}{\theta}

\newcommand{\mech}{\mathcal{M}}

\newcommand{\system}{\textsc{DP-Voxlet}\xspace}

\renewcommand{\paragraph}[1]{\vspace*{2pt} \noindent \textbf{#1}}

\title{\system: Provable Speaker Anonymization for Disentangled Speech Representations}

\author[affiliation={1}]{Ivoline}{Ngong}
\author[affiliation={1}]{Jack}{D'Iorio}
\author[affiliation={1}]{Hailey}{Schoppe}
\author[affiliation={2}]{Christopher}{Liberatore}
\author[affiliation={2}]{Nichole}{Schimanski}
\author[affiliation={2}]{Taisa}{Kushner}
\author[affiliation={1}]{Joseph P.}{Near}

\address{
    $^1$ University of Vermont, USA\\
    $^2$ Galois, Inc., USA
}

\email{kngongiv@uvm.edu, Jack.DIorio@uvm.edu, Hailey.Schoppe@uvm.edu, cliberatore@galois.com, nls@galois.com, taisa@galois.com, jnear@uvm.edu}

\keywords{privacy, anonymization, differential privacy}

\begin{document}
\maketitle

\begin{abstract}
  Systems for speaker anonymization obfuscate the speaker of an utterance, while maintaining its original semantic contents and prosody. Recent solutions for speaker anonymization rely on learned representations that disentangle an utterance into semantic contents and speaker properties. To anonymize an utterance, these systems replace the speaker properties while leaving the semantic contents unchanged---an approach that can produce strong results on empirical measures of privacy.

  In this work, we introduce speaker differential privacy, a formal definition of speaker anonymization based on the framework of differential privacy, and a mechanism for speaker anonymization that provably satisfies the definition. In contrast to prior heuristic-based anonymization systems, our approach enables a provable lower bound on re-identification success rate (e.g. equal error rate) for any possible adversary. We implement our approach in a framework that is compatible with existing disentangled representations. Compared to the prior work on differential privacy for speaker anonymization, our approach achieves significantly higher utility.
\end{abstract}

\section{Introduction}

Systems for \emph{speaker anonymization} attempt to obfuscate the speaker of an input utterance, while maintaining the original semantic contents and prosody of the input. Many recent solutions for speaker anonymization (e.g.~\cite{qin2023openvoice, ju2024naturalspeech, guo2024vec2wav, chen2022controlvc}) rely on \emph{disentangled representations}: learned representations of an utterance that split the information into pieces representing the semantic contents of the utterance and the properties of the speaker, such that mutual information between the two is minimized.

To achieve speech privacy, these systems leave the semantic contents of the utterance unchanged, and modify the representation of the speaker to make the output sound like a different person. For example, the team that achieved the best anonymization results in the 2024 Voice Privacy Challenge 
disentangled the input utterance into semantic, emotional, and speaker information, and then replaces the speaker information with a randomly-selected speaker from a predetermined pool. Existing systems perform well on empirical benchmarks, but do not provide provable guarantees of privacy.

This work introduces \emph{speaker differential privacy}, a formal definition of speaker anonymization based on differential privacy~\cite{dwork2006calibrating, dwork2014algorithmic}, and \system, a general framework for achieving speaker differential privacy using existing approaches for disentangled representation of speech. Speaker differential privacy provides theoretical bounds on the ability of \emph{any} adversary to accurately guess the original speaker of an anonymized utterance, providing a provable guarantee independent of any specific empirical benchmark.
\system works by applying a random perturbation to the disentangled speaker representation, rather than choosing the new speaker from a predefined pool. The randomized perturbation is carefully designed to ensure differential privacy for the resulting utterance.

Existing solutions for speech disentanglement are not designed for random perturbation of speaker information, and perform poorly when the speaker embedding is perturbed. To solve this problem, we develop a differentially private variational autoencoder architecture to perturb the speaker embedding while remaining in the well-conditioned portion of the embedding space. An empirical devaluation of \system using the 2024 Voice Privacy Challenge benchmark shows that our approach is competitive with state-of-the-art anonymization systems on empirical benchmarks, while also providing provable guarantees.

\paragraph{Contributions.} In summary, our contributions are:
\begin{itemize}[topsep=0pt, itemsep=0pt, leftmargin=10pt]
\item We define \emph{speaker differential privacy}, a formal privacy definition for speaker anonymization in disentangled speech representations.
\item We describe a perturbation mechanism for speaker privacy and prove it satisfies the definition.
\item We develop \system, a framework for applying speaker differential privacy in existing voice conversion systems based on a differentially private variational autoencoder.
\item Through empirical evaluation, we show that \system is competitive with existing anonymization solutions that do \emph{not} provide provable guarantees.
\end{itemize}

\subsection{Background: Differential Privacy}

In the area of data privacy, \emph{differential privacy} (DP)~\cite{dwork2006calibrating,dwork2014algorithmic} has emerged as the standard framework for controlling privacy leakage when releasing data. It provides provable guarantees limiting the ability of an adversary to learn new information about individuals in the data, even when the adversary has access to auxiliary information.  In DP, a {mechanism} $\mech$ refers to a randomized algorithm that takes as input a database $x$ and releases some statistics $\mech(x)$ if of the database in some abstract output space. DP requires that it be hard to distinguish between pairs of neighboring databases, $S$ and $S'$ based upon the data provided from $\mech(\cdot)$, where two databases are considered neighboring if they differ in one person's data.

Formally, a {mechanism} $\mech$ satisfies $(\epsilon, \delta)$-DP if for all neighboring databases $S$ and $S'$, and for all possible sets of outcomes $O$, $Pr[\mech(S) \in O] \leq e^\epsilon \Pr[\mech(S') \in O] + \delta$.

Differentially private mechanisms typically use random perturbations to achieve this definition. For example, the \emph{Gaussian mechanism} adds Gaussian-distributed random noise to the result of a database query. Our work is based on a variant of differential privacy called \emph{Gaussian differential privacy}~\cite{dong2022gaussian} that directly bounds the tradeoff between true- and false-positive rates for an adversary attempting to guess properties of an individual.

\subsection{Related Work}

Signal processing methods for speaker anonymization~\cite{patino2020speaker, vaidya2019you, srivastava2020evaluating, qian2017voicemask} are generally simple and do not require any training, but degrade speech significantly when used to produce high levels of privacy. More recent work based on voice conversion models has begun to explore disentangled representations~\cite{yoo2020speaker, fang2019speaker, prajapati2022voice}. Generally, these approaches work by extracting an x-vector (containing speaker properties) from an input utterance using an encoder, replacing the x-vector with a random one from a pool of speakers~\cite{srivastava2020design}, and generating a new utterance using a decoder. Recent work suggests that using discrete tokens to represent semantic content can improve the disentangled representation and avoid speaker identity information leaking into the semantic content~\cite{champion2022disentangled, champion2023anonymizing}. Our approach leverages this prior work on disentanglement, adding a provable guarantee.

The closest work to ours is that of Shamsabadi et al.~\cite{shamsabadi2023differentially}. Their approach also aims to ensure differential privacy for the speaker of an utterance. In contrast to our approach, they randomly perturb \emph{all} features of the input utterance, which reduces utility by perturbing the semantic contents of the utterance. As a result, their solution requires extremely large privacy budgets to achieve good utility.
Also inspired by differential privacy, Tran and Soleymani~\cite{tran2023speech} selectively perturb specific parts of a disentangled representation, guided by a learned \emph{saliency estimator}. Their approach does not provide a formal guarantee and reduces utility.
Han et al.~\cite{han2020voice} introduce \emph{voice indistinguishability}, a formal definition also based on differential privacy. However, their definition is based on metric differential privacy~\cite{chatzikokolakis2013broadening}, so privacy protection degrades with distance between speakers.

\section{Formal Speaker Anonymization}

Our goal is to define formal, compositional guarantees of speaker privacy that have a natural real-world interpretation.
In our setting, we want to limit the adversary's ability to distinguish which of two possible speakers produced the input to the mechanism, based on the  output of the anonymization system.

\subsection{System Model, Assumptions, and Limitations}

We assume the existence of an encoder $\encode(x) = (S, C)$ where $S$ denotes the speaker's attributes (as a \emph{speaker embedding}) and $C$ represents the semantic contents of the utterance $x$. We also assume the existence of a decoder $\decode(S, C) = x$ that decodes these pieces into a waveform. We assume that the training process assures that $\forall x$, $\decode(\encode(x)) = x$ to some arbitrary accuracy. 

We further assume that the encoder has been trained to ensure that the utterance content $C$ \emph{does not include any information about the identity of the speaker}. 
Formally, consider any two speakers with utterances $x$ and $x'$, where the two utterances express the same contents. If $\encode(x) = (S, C)$ and $\encode(x') = (S', C')$, then $C = C'$. This separation of the components of the latent representation is vital for our privacy guarantees; if there is leakage of the speaker's attributes into $C$ then an adversary may be able to identify the original speaker with high confidence.

One way to formalize this assumption is in terms of true and false positive rates for an adversary:
\begin{assumption}
Consider any utterance $x$ spoken by speaker $s$ such that $\encode(x) = (S, C)$, and an adversary $\mathcal{A}(s, C)$ that attempts to determine whether the semantic contents $C$ of an utterance were spoken by speaker $s$. For a well-disentangled encoder $\encode$, no adversary exists which performs better than chance at this task.
\label{asm:disentangle}
\end{assumption}

\paragraph{Limitations.}
Assumption~\ref{asm:disentangle} is impossible to prove.
In practice, the definition of $\encode$ will be in terms of trained models that produce $S$ and $C$; prior work has demonstrated disentanglement empirically~\cite{chen2022controlvc, guo2024vec2wav, ju2024naturalspeech, qin2023openvoice}, but it still may be possible for speaker information to leak into the contents $C$. When disentanglement is imperfect, our approach does not ensure {perfect} speaker privacy, because as speaker information may leak into $C$. 
However, our empirical results suggest that this leakage is minimal in practice.

\subsection{Privacy Definition}

Our privacy definition considers two speakers $s_1$ and $s_2$, and limits the ability of an adversary to determine which of the two spoke an anonymized utterance.
It is patterned on the definition of Gaussian differential privacy~\cite{dong2022gaussian}, which defines privacy in terms of \emph{trade-off functions}. A trade-off function characterizes the optimal relationship over a rejection rule $0\leq R \leq 1$, which rejects a null hypothesis test that the data exposed by $\mech$ came from dataset $S$ and not $S'$. This optimality is derived from the relationship between type-I errors (false positives, $\alpha_{R} = \mathbb{E}_P[R]$) and type-II errors (false negatives, $\beta_{R} = \mathbb{E}_Q[R]$) that an optimal adversary will make when observing outputs of a mechanism. The trade-off $T$ on two probability distributions $P$ and $Q$ characterizes the trade-off between these two types of errors in distinguishing between draws from $P$ and $Q$.
\begin{definition}[Trade-off function~\cite{dong2022gaussian}]
\label{def:trade-off}
  For any two probability distributions $P$ and $Q$ on the same space, the trade-off function $T(P, Q): [0, 1] \rightarrow [0, 1]$ is defined as:
  \[T(P, Q)(\alpha) = \inf \{ \beta_R : \alpha_R \leq \alpha \} \]
  where the infimum is taken over all rejection rules R.
\end{definition}

The formal definition also includes a parameter $\mu$ (corresponding to the privacy parameter $\mu$ in Gaussian differential privacy~\cite{dong2022gaussian}) that tunes the strength of the guarantee. Setting $\mu=0$ results in perfect privacy under this definition---an adversary who observes an output of the system will believe that every possible speaker is equally likely to be the original speaker. This kind of perfect privacy is usually not achievable; instead, we set $\mu > 0$ (but ensure that $\mu$ remains small) to ensure strong privacy.

\begin{definition}[Gaussian differential privacy~\cite{dong2022gaussian}]
\label{def:gaussian-diff-privacy}
A mechanism $\mathcal{M}$ is said to satisfy $\mu$-Gaussian Differential Privacy ($\mu$-GDP) if it
is $G_\mu$-DP. That is,
\[T(\mathcal{M}(S), \mathcal{M}(S')) \geq G_\mu \]
for all neighboring datasets S and S', where $G_\mu$ is defined as:
\[G_\mu(\alpha) = \Phi(\Phi^{-1}(1 - \alpha) - \mu)\]
and $\Phi$ denotes the standard normal CDF.
\end{definition}

Our definition of speaker privacy extends this idea to utterances and the encoder $\encode$ and decoder $\decode$.

\begin{definition}[Speaker differential privacy]
  A privacy mechanism $\mech$ satisfies \emph{$\mu$-speaker differential privacy} if for all utterances $x_1$ spoken by $S_1$ and $x_2$ spoken by $S_2$ such that $\encode(x_1) = (S_1, C)$ and $\encode(x_2) = (S_2, C)$, then:
  \[ T(\decode(\mech(\encode(x_1))), \decode(\mech(\encode(x_2)))) \geq G_\mu \]
\end{definition}

\subsection{Privacy Mechanism}

Satisfying our definition is difficult, because it quantifies over both known and unknown speakers. As a result, heuristic approaches such as selecting randomly from a set of known speakers does not satisfy the definition. Instead, inspired by DP literature, we sample the latent speaker embedding space, leveraging the idea of \emph{$L_2$ global sensitivity}~\cite{dwork2014algorithmic}.

We first bound the maximum $L_2$ distance between embeddings of any two speakers $S_1$ and $S_2$ in the latent space:$\Delta_s = \max_{S_1, S_2} \lVert S_1 - S_2 \rVert_2$.  We ensure bounded $L_2$ global sensitivity by \emph{clipping}: we enforce an upper bound on the $L_2$ norm of the speaker embedding by scaling the embedding to have a magnitude no larger than a constant $U$. The perturbation Gaussian noise level, $\sigma$, is computed as the ratio between the maximum $L_2$ distance $U$ and the desired privacy level $\mu$ (so $\sigma$ is inversely proportional to $\mu$). Effectively, this noise addition is a method for \emph{generating a new, hypothetical speaker} who is not a real person. If enough noise is used, it will be difficult for an adversary to determine the original speaker with high confidence. 
The Gaussian speaker mechanism algorithm is shown in Algorithm \ref{alg:gaussian_mechanism}; it satisfies $\mu$-speaker differential privacy (proof available in the code repository).

\begin{theorem}
The Gaussian speaker mechanism $\mech_G$ satisfies $\mu$-membership speaker privacy.
\end{theorem}

\begin{algorithm}[t]
\caption{Gaussian speaker mechanism $\mech_G$}
\label{alg:gaussian_mechanism}
\begin{algorithmic}
\Require speaker embedding vector $S$ from $\encode(x) = (S,C)$
\Require maximum $L_2$ perturbation, $U$
\Require privacy parameter, $\mu$
\State $S' \leftarrow \textit{$L_2$clip}(S,U)$
\State $\sigma = \frac{U}{\mu}$
\State $\hat{S} = S' + \mathcal{N}(\sigma^2)$
\Ensure $(\hat{S}, C)$
\end{algorithmic}
\end{algorithm}

\subsection{Provable Adversarial Lower Bounds}

The Gaussian speaker mechanism and the definition of speaker differential privacy enables a theoretical lower bound on the success rate of an adversary who attempts to re-identify the speaker of an anonymized utterance. The definition of speaker differential privacy is in terms of a trade-off function between false positives and false negatives for \emph{any} adversary attempting a distinguishing task.

\begin{figure}
  \centering
  \includegraphics[width=.49\textwidth]{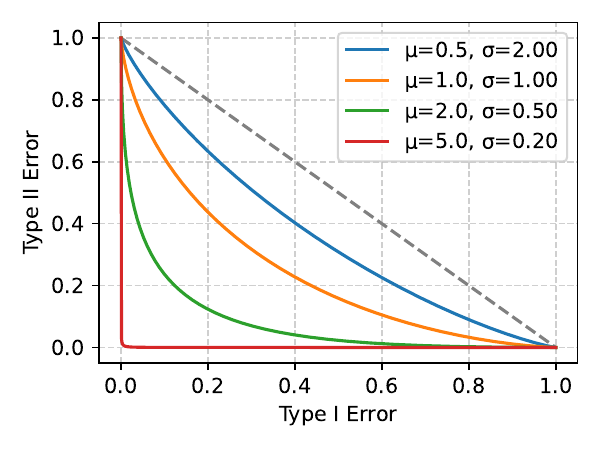}
  \caption{EER Lower Bound, computed from Defs. \ref{def:trade-off} and \ref{def:gaussian-diff-privacy}.}
  \label{fig:eer_bound}
\end{figure}

Viewed through this lens, the Gaussian speaker mechanism ensures a lower bound on equal error rate (EER) for \emph{any adversary} based on the variance of the Gaussian noise added by the mechanism. Figure~\ref{fig:eer_bound} illustrates the trade-off function for the Gaussian speaker mechanism at various settings of the privacy parameter $\mu$. The point where type-I and type-II error are equal is called the equal error rate (EER), and it occurs on the diagonal $y = x$ since the errors are symmetric. As the privacy parameter $\mu$ gets closer to zero, the lower bound on EER improves (to ``perfect privacy''---an EER of 50\%---when $\mu=0$). For common settings like $\mu=1$, the lower bound on EER is around 35\%. Note that because of the relationship between $\mu$ and $\sigma$ in Algorithm \ref{alg:gaussian_mechanism}, when $U = 1$, $\mu = \sigma^{-1}$.

\section{The \system System}

\begin{figure*}
  \centering
  \includegraphics[trim={0 0.3in 0 0},width=.99\textwidth]{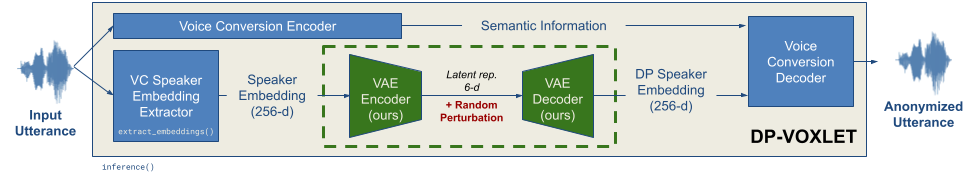}
  \caption{Architecture of \system}
  \label{fig:system}
\end{figure*}

We implement \system, a framework for adding formal speaker anonymization to an existing voice conversion system. Figure~\ref{fig:system} summarizes the architecture of \system. The framework requires the voice conversion system to be organized as an encoder and decoder, and to separate speaker-independent semantic information about an utterance from the speaker-identifying information (e.g. speaker embedding). The \system framework implements the required perturbation for the speaker-identifying information to ensure membership speaker privacy.
Our implementation is open-source.\footnote{\scriptsize \url{https://github.com/uvm-plaid/dpvc}}

\paragraph{Compatibility with existing encoder/decoder systems.}
Many recent voice conversion systems are structured this way, and can be easily integrated with \system by writing a small wrapper class to expose the components of the system. \system is implemented as a Python library using PyTorch, and integrates with voice conversion systems that also use PyTorch. Our current implementation includes wrappers for OpenVoice~\cite{qin2023openvoice}, NaturalSpeech3~\cite{ju2024naturalspeech}, vec2wav2.0~\cite{guo2024vec2wav}, and ControlVC~\cite{chen2022controlvc}. Writing a new wrapper requires implementing two methods: \texttt{extract\_embedding}, which extracts a speaker embedding from an input audio file, and \texttt{inference} (both noted in Figure \ref{fig:system}), which performs voice conversion on an input audio file given a target speaker embedding.

\paragraph{Synthesizing good speaker embeddings.}
VC systems typically represent the speaker embedding using a 192- or 256-dimensional vector. Perturbing this vector directly with the Gaussian speaker mechanism satisfies membership speaker privacy, but results in low-quality speech output because voice conversion decoders typically do not generalize well over the entire space of speaker embeddings. Instead, these decoders work well only for regions of the embedding space characterized well by their training data, and produce garbled output for embeddings outside these regions. Applying the Gaussian speaker mechanism directly often results in an embedding outside a well-behaved region, and thus unintelligible output.

To address this challenge, we train a variational autoencoder (VAE)~\cite{kingma2013auto} to represent the ``valid'' speaker embeddings for a given voice conversion system using a low-dimensional embedding. We train the VAE on speaker embeddings extracted from the Common Voice dataset~\cite{ardila2020common} using the target voice conversion system. The strong regularization of the VAE objective means that the low-dimensional embedding is robust to perturbations, since the VAE's decoder is trained to map any low-dimensional representation to a ``valid'' speaker embedding. \system applies the Gaussian speaker mechanism to the low-dimensional VAE embedding, then uses the VAE's decoder to construct the final speaker embedding for voice conversion.

\paragraph{Mitigating large perturbations.}
Large perturbations to the speaker embedding can produce significantly degraded speech, despite the use of the VAE. To mitigate this effect, we clamp each dimension of the latent representation to have bounded absolute value (i.e. we enforce a limit on the $L_\infty$-norm of the latent representation). This clamping ensures that the perturbed latent representation will lie in the region that the VAE's decoder ``understands,'' so that the decoder will produce a valid speaker embedding. This kind of clamping neither improves nor degrades privacy (since differential privacy is closed under post-processing), but it can avoid garbled output due to large perturbations.

\section{Evaluation}

To evaluate the performance of \system empirically, we used the 2024 Voice Privacy Challenge benchmark~\cite{tomashenko2024voiceprivacy}, which evaluates both privacy and utility. The benchmark evaluates privacy using a speaker verification model: it anonymizes two utterances (from the same speaker or two different speakers) and then asks the speaker verification model whether the anonymized utterances come from the same speaker. We used the semi-informed attacker, where the speaker verification model is trained on anonymized speech. The benchmark measures privacy using equal error rate (EER): the value for which the speaker verification model's false positive and false negative rates are equal. An EER of 50\% indicates perfect privacy. The benchmark measures utility using an automated speech recognition model (ASR), via word error rate (WER). A WER of 0\% indicates perfectly intelligible speech. The benchmark uses the test split of the librispeech dataset~\cite{panayotov2015librispeech} to perform the experiment and the OpenVoice~\cite{qin2023openvoice} system as the VC base method.

\begin{figure}
  \centering
  \includegraphics[width=.5\textwidth]{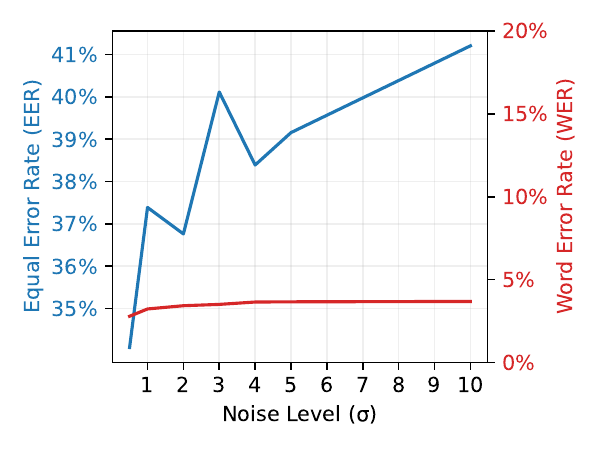}
  \newcommand{\rs}{\hspace*{-3pt}}
\vspace*{-5pt}
    \begin{tabular}{|l|llllll|}
      \hline
     & $\sigma \mathord{=} 0$\rs & $\sigma \mathord{=} .5$\rs & $\sigma \mathord{=} 1$\rs & $\sigma \mathord{=} 2$\rs & $\sigma \mathord{=} 5$\rs & $\sigma \mathord{=} 10$\rs\\
     & $\mu \mathord{=} \infty$\rs & $\mu \mathord{=} 2$\rs & $\mu \mathord{=} 1$\rs & $\mu \mathord{=} .5$\rs & $\mu \mathord{=} .2$\rs & $\mu \mathord{=} .1$\rs\\
    \hline
    \rs EER \rs& 4.6\%\rs & 34.0\%\rs & 37.3\%\rs & 36.8\%\rs & 39.2\%\rs & 41.2\%\rs\\
    \rs WER \rs& 3.0\%\rs & 2.8\%\rs & 3.1\%\rs & 3.3\%\rs & 3.5\%\rs & 3.4\%\rs\\
    \hline
  \end{tabular}

  \caption{Experimental Results: Equal Error Rate (EER) and Word Error Rate (WER) as calculated by the 2024 Voice Privacy Challenge benchmark on the librispeech-test dataset, using \system with the OpenVoice system.}
  \label{fig:experiment_graph}
\end{figure}

Our results appear in Figure~\ref{fig:experiment_graph}. As the variance of the perturbation rises (i.e., noise level $\sigma$, X-axis), the level of measured privacy as indicated by EER also rises (left Y-axis). WER (right Y-axis) also rises slightly with perturbation variance, but it remains below 4\%. Figure~\ref{fig:experiment_graph} is the result of a single random perturbation of each evaluation setting across the test dataset; the variance is due to randomness in the anonymization process.

\paragraph{Comparison to Voice Privacy Challenge submissions.}
Our results are competitive with many of the submitted anonymization systems for the 2024 Voice Privacy Challenge,
even though none of those systems provided provable guarantees. Out of 36 submissions to the challenge, only 6 achieved an EER higher than 40\%; in our experiments, \system's highest EER was 41.2\%.

\paragraph{Comparison to \cite{shamsabadi2023differentially}.}
Our approach produces significantly improved EER compared to the closest related work, Shamsabadi et al.~\cite{shamsabadi2023differentially}. Their results, also on librispeech (but with a different speaker verification model) show a maximum EER of less than 20\%, and require much larger privacy budgets.

\section{Conclusion}

We have introduced speaker differential privacy, a formal definition of speaker anonymization for disentangled representations inspired by differential privacy that enables provable lower bounds on adversary success in re-identifying speakers in anonymized utterances. We describe a mechanism for this definition and implement it in \system. Our empirical results suggest that our approach provides robust privacy protection while maintaining high-quality output speech.

\section{Acknowledgments}
This work was supported by the Intelligence Advanced Research Projects Activity (IARPA) via Department of Interior/Interior Business Center (DOI/IBC) contract number 140D0424C0066. The U.S. Government is authorized to reproduce and distribute reprints for Governmental purposes notwithstanding any copyright annotation thereon. The views and conclusions contained herein are those of the authors and should not be interpreted as necessarily representing the official policies or endorsements, either expressed or implied, of IARPA, DOI/IBC, or the U.S. Government.

\bibliographystyle{plain}
\bibliography{refs, speech_refs}

\appendix

\end{document}